\documentclass[sigconf]{acmart}

\pdfoutput=1

\usepackage{acronym}
\acrodef{ALL}{Activity Led Learning}
\acrodef{CEM}{Computing, Electronics \& Mathematics}
\acrodef{EEC}{Engineering, Environment \& Computing}
\acrodef{EU}{European Union}
\acrodef{GDPR}{General Data Protection Regulation}
\acrodef{IDE}{Integrated Development Environment}
\acrodef{ILO}{Intended Learning Outcome}
\acrodef{MCQ}{Multiple Choice Questionnaire}
\acrodef{MEQ}{Module Evaluation Questionnaire}
\acrodef{SFINAE}{Substitution Failure Is Not An Error}
\acrodef{VLE}{Virtual Learning Environment}
\acrodef{VM}{Virtual Machine}

\usepackage[super]{nth} 
\usepackage{numprint}

\usepackage{tikz}
\usepackage{pgfplots}
\pgfplotsset{width=8cm,compat=1.9}
\usetikzlibrary{pgfplots.dateplot}
\usepackage{pgfplotstable}

\usepackage{filecontents}

\definecolor{cb1}{rgb}{0.00,0.00,0.00} 
\definecolor{cb2}{rgb}{0.90,0.60,0.00} 
\definecolor{cb3}{rgb}{0.35,0.70,0.90} 
\definecolor{cb4}{rgb}{0.00,0.60,0.50} 
\definecolor{cb5}{rgb}{0.80,0.40,0.00} 
\definecolor{cb6}{rgb}{0.00,0.45,0.70} 
\definecolor{cb7}{rgb}{0.80,0.60,0.70} 
\definecolor{cb8}{rgb}{0.95,0.90,0.25} 
\definecolor{cb9}{rgb}{0.5,0.5,0.5} 

\colorlet{ptavg}{cb4}
\colorlet{pt1}{cb7}
\colorlet{pt2}{cb6}

\usepackage{booktabs} 

\setcopyright{rightsretained}

\acmDOI{10.475/123_4}

\acmISBN{123-4567-24-567/08/06}

\copyrightyear{2019} 
\acmYear{2019} 
\setcopyright{acmcopyright}
\acmConference[CEP '19]{Computing Education Practice}{January 9, 2019}{Durham, United Kingdom}
\acmBooktitle{Computing Education Practice (CEP '19), January 9, 2019, Durham, United Kingdom}
\acmPrice{15.00}
\acmDOI{10.1145/3294016.3294018}
\acmISBN{978-1-4503-6631-1/19/01}

\begin{document}
\title{Computing with Codio at Coventry University}
\subtitle{Online virtual Linux boxes and automated formative feedback}

\author{David Croft}
\affiliation{%
  \institution{Coventry University}
  \streetaddress{}
  \city{Coventry}
  \state{U.K.}
}
\email{David.Croft@coventry.ac.uk}

\author{Matthew England}
\affiliation{%
  \institution{Coventry University}
  \streetaddress{}
  \city{Coventry}
  \state{U.K.}
}
\email{Matthew.England@coventry.ac.uk}

\renewcommand{\shortauthors}{D. Croft and M. England}

\begin{abstract}
We describe our experience using Codio at Coventry University in our undergraduate programming curriculum.  Codio provides students with online virtual Linux boxes, and allows staff to equip these with guides written in markdown and supplemental tasks that provide automated feedback.  The use of Codio has coincided with a steady increase in student performance and satisfaction as well as far greater data on student engagement and performance.
\end{abstract}

%
%

\ccsdesc[500]{Social and professional topics~Computing education}
\ccsdesc[500]{Applied computing~Interactive learning environments}
\ccsdesc[500]{Applied computing~Computer-assisted instruction}
\ccsdesc[500]{Applied computing~Computer-managed instruction}
\ccsdesc[500]{Applied computing~Learning management systems}

\keywords{Programming Education; Virtual Machines; Automated Feedback}

\maketitle

\section{Introduction}

The School of \ac{CEM} within the Faculty of \ac{EEC} at Coventry University runs 7 BSc degrees which share initial modules on programming led by the authors.  This includes BSc Computer Science, but also degrees split with other fields such as BSc Information Technology for Business, and BSc Mathematics \& Data Science.  
We describe recent innovations in these programming classes.

\subsection{Situation Prior to Codio}


Prior to the initiatives we discuss in this paper, the content in these programming classes was fairly traditional\footnote{
We note the presence of innovative practice elsewhere in the computing curriculum however, in particular our \ac{ALL} projects \cite{CovALL-CC}.  Briefly:  all students start their degrees with an integrative project that requires ideas from all the modules they are studying but is led by the problem statement and can be developed differently depending on the students interests.  The ALL initiative began prior to the ones we focus on here and continues to this date.  The ALL classes offer students an opportunity to put their modules into context, and practice their programming skills as part of a larger project.  This paper focuses on how we have improved the actual programming classes which take place alongside ALL and provide those skills.
}.  There were a mix of lectures based around slides and labs based around a print out (or more recently a pdf on screen) with a series of problems to tackle. 
Students attempt these, getting help from staff when needed, with solutions released at some later date.  To gain formative feedback students had to do one of the following:
\begin{itemize}
    \item Wait until the nominated release date of solutions and check their answers.  This is usually a week later by which time the student may have forgotten the details of the exercise and their solution, dampening the learning that can take place.
    \item Request feedback from staff in lab sessions.  Many students seem to find this difficult, particularly in Year 1: perhaps over fear/embarrassment of asking for help, or simply not reaching a stage where feedback can be useful during labs.
\end{itemize}
Finding a method to more effectively deliver meaningful feedback to student was a key goal for the authors.  This motivation was compounded by a rapid increase in student numbers in these classes.  These have doubled over just a few years to reach $~480$ Stage 1 students taking programming in the 2017/18 academic year.  
The corresponding increase in staff is not of the same scale, putting a greater strain on the delivery of formative feedback.  

Even where staff resources do rise we have a related problem of consistency of feedback.  For example, in Semester 1 2017/18 the first year programming module took place over 40 hours in 20 different lab session taught by a combination of 12 different staff members.  Even with the best of intentions, consistency of quantity, quality and content of formative feedback was not possible.
We decided to use automated testing to improve the situation.  Our aims were:
\begin{enumerate}
    \item more interactive lab material with a degree of gamification;
    \item instant formative feedback for students; and 
    \item better monitoring of student progress on modules. 
\end{enumerate}

\subsection{Codio}

The decision was made to implement our solutions via Codio\footnote{https://codio.co.uk/}, a commercial cloud based infrastructure focused on STEM, and in particular Computer Science, education.  

Teachers prepare Codio Units for students.  Each unit is built around a Linux based \ac{VM} environment\footnote{Each student has their own VM - the actions of one student do not affect the others.}, augmented with a guide (written by the tutors) which can contain a variety of tasks to provide instant formative feedback.  

In Section \ref{SSEC:IDE} we describe the benefits we found from using Codio as a development environment and in Section \ref{SSEC:Feedback} we focus on the most important one: the automated programming feedback now provided at Coventry.  In Section \ref{SEC:Value} we describe out initial findings on the value this had added to our students education, in Section \ref{SEC:Challenges} we summarise some challenges we encountered, and we finish in Section \ref{SEC:Future} with some plans for the future.

\section{Our Experiences with Codio}

\subsection{Development environment}
\label{SSEC:IDE}

One of the more irritating factors in teaching programming are the limitations imposed by developing on a centrally managed IT system.  
Programming classes at Coventry are taught in PC labs, and whilst some students may bring personal devices, most use those machines in class.
These machines are restricted in a number of ways for obvious security and management issues: for example, students are not able to install additional software on them.  This  extends to code libraries and \acp{IDE}.  Tutors must decide over the summer what software is required for all classes in the year ahead.  Even when this level of advance planning is achieved it shuts down the ability for students to research and choose their own libraries for projects and extra curricular work.

At Coventry the initial approach to resolving this issue was the use of cloud \ac{VM} providers on an ad-hoc basis in individual teaching modules.  The passing of the \ac{GDPR} in 2016 put an end to this practice as it left the university open to significant fines\footnote{For details on the GDPR and practical examples see for example \cite{VB17}.}.
The decision was made to use a single \ac{GDPR} compliant cloud solution.  Although Codio does not have as many development features as other providers, its focus on education and testing functionality as detailed below, made it the best choice for our teaching purposes.

\begin{figure*}[t]
    \centering
    \includegraphics[width=\textwidth]{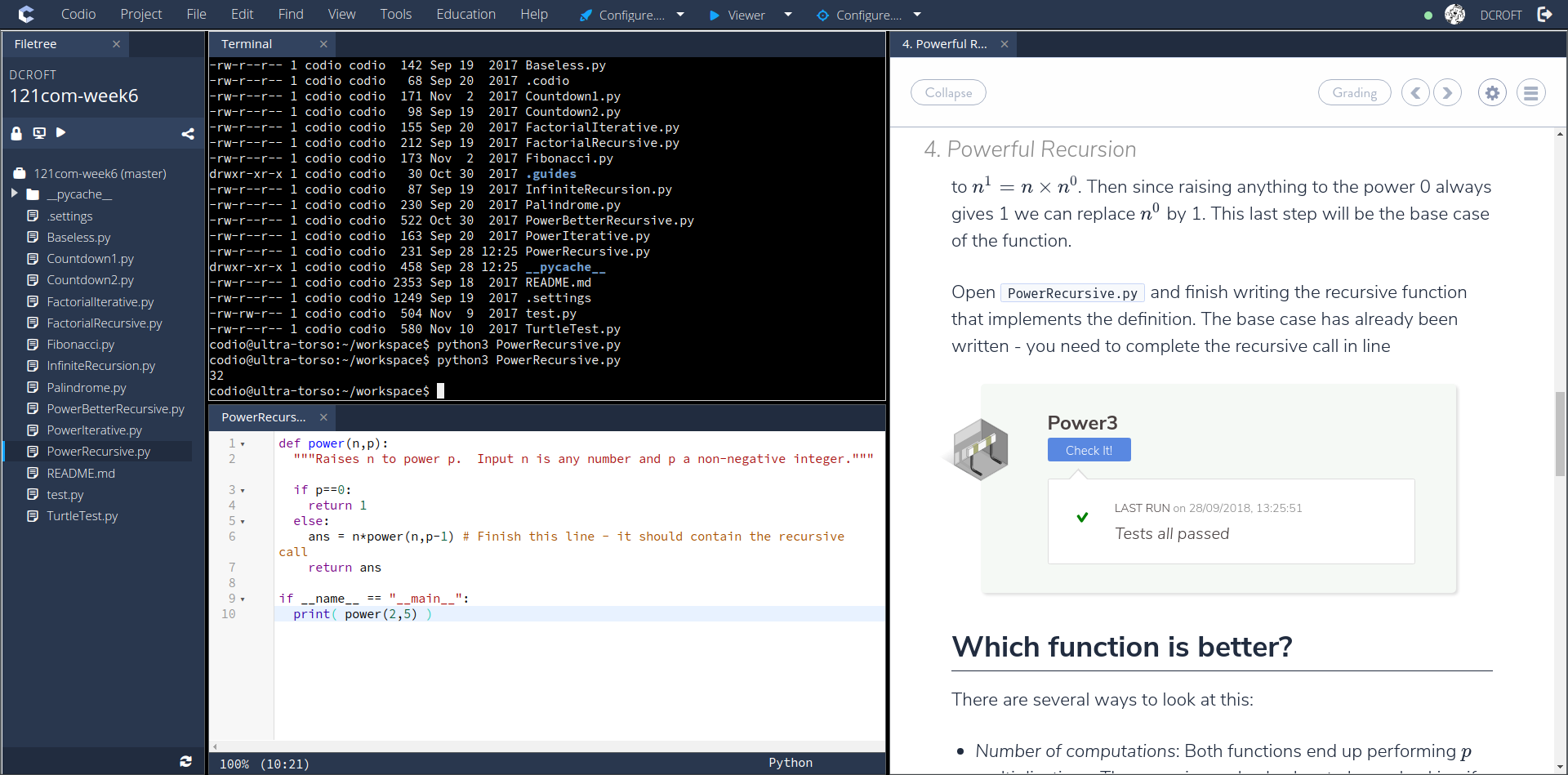}
    \caption{The Codio environment}
    \label{fig:codio}
\end{figure*}

Figure \ref{fig:codio} shows a screenshot from a typical Codio session:  on the left we have the file tree of the main workspace; middle bottom a text editor\footnote{The editor has basic IDE tools such as syntax highlighting and auto-complete of existing function and variable names.}; middle top a Linux bash terminal where students run and test their code themselves; and on the right the guide.  The guide is written by the teacher in \texttt{Markdown} with simple environments for text, mathematics (using \LaTeX) and code (with syntax highlighting for common languages).  Most importantly, the guides contain embedded assessments.  The assessments can be multiple choice, fill in the blanks etc. but of most interest are those that run a test script prepared by the teachers and hidden to the students, as discussed in Section \ref{SSEC:Feedback}.

As an environment Codio offers two further notable advantages.  First, it essentially doubles as a cloud storage system for a student's programming work: they can access Codio from anywhere they have internet and pickup their work with the code and files where they left of, removing the need to backup onto USB drives or email files at the end of each class\footnote{It also removes one of the key excuses for not doing homework!}.
Secondly, if students are having issues then staff are able to remotely open their projects and look into the issue directly.  This saves considerable time compared to lengthy email exchanges that contain code snippets without syntax highlighting (and often with email software removing indentation), which may not actually even contain the bug.

\subsection{Automated feedback on code}
\label{SSEC:Feedback}

\begin{figure}[t]
\includegraphics[trim={1.8cm 12.95cm 11cm 13.1cm},clip]{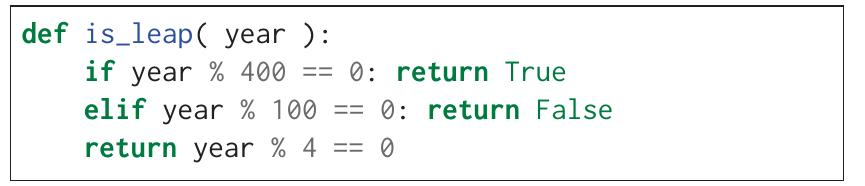}
    \caption{Example of solution expected from students.}
   \label{fig:code}
\end{figure}

At Coventry we still have a significant proportion of students who have not programmed much before university\footnote{While the UK government may now mandate programming education in schools this has not fully filtered through to UG entrance yet, and even then does not take into account our substantial international student base.} and so there is a need for regular and detailed support on often fairly basic points.  Even amongst those who have programmed before, effective software testing is a separate skill that few students enter with.

We thus implemented our own tests that would judge a student's code.  This offers a number of advantages:
\begin{itemize}
    \item Our tests are usually more thorough than those a student would implement.
    \item A student gets validation that their code works, even if it differs from the \emph{model solution} provided by the teacher.
    \item This feedback is available at any time to suit the students, in or outside class.  They are able to enter a cycle of development and testing without waiting to see the teacher.  
    \item While not contributing to their grade, some student derive satisfaction from  a \emph{high score} on a unit (gamification).
\end{itemize}
The tests and feedback can take any form (that may be delivered by terminal) as they are coded by the instructor.  For us, they are usually a detailed set of unit and/or system tests, although in some cases we can pose further restrictions on students if appropriate for the topic of study (e.g. no use of loops on the tasks for studying recursion).  We usually let students see which test cases fail, and sometimes augment this with additional text feedback that expands on the error message from the language or points out common mistakes (e.g. if their code produces a Python \texttt{IndexError} we might remind them that Python lists index from $0$).

\subsubsection*{Example}

An early lab task has students writing a function to identify if a given positive integer is a leap year or not\footnote{Without using the language's \texttt{datetime} libraries - something tests can easily detect.}.  The complete leap year logic is shown in Python in Figure \ref{fig:code}.
The task is partially intended to demonstrate a student's ability to correctly declare and call a function.  The main aim is to provoke a discussion regarding the importance of correct requirements gathering and to demonstrate the dangers of making unsupported assumptions when writing their code.

The majority of students will initially write functions that simply test if a year is divisible by 4 and will test only a handful of values based on that condition (usually the current and adjacent years) while our automated tests would examine all 4 branches of the logic.
The end result is a demonstration of good development practice as well as ensuring that all students have completed a task that tests their ability to write functions that test multiple logical conditions.

\section{Value added}
\label{SEC:Value}

Codio was initially trialled on a single module basis in 2016/17 with a roll out to all first year programming in 2017/18.  Going into the 2018/19 academic year Codio use has expanded into further modules and courses.

\subsection{Monitoring and Engagement}

Codio provides monitoring features at both class and individual level.  
We have developed additional functionality to record historical engagement and compare with performance on summative assessment (which takes place outside of Codio).  We aim to use this data to improve engagement, and hence performance and retention.  

Our main goal is for students to gain a functional level of programming competence, 
and a pre-requisite for this is simply putting in the time to practice.  The often quoted figures are ``\emph{10 years or \numprint{10,000} hours to become an expert}'' \cite{johnhayes1989, dr.benjaminbloom1985}. 
Of course, for that time to be effective students should be challenged to explore more complex programs rather than repeating the same tasks.  

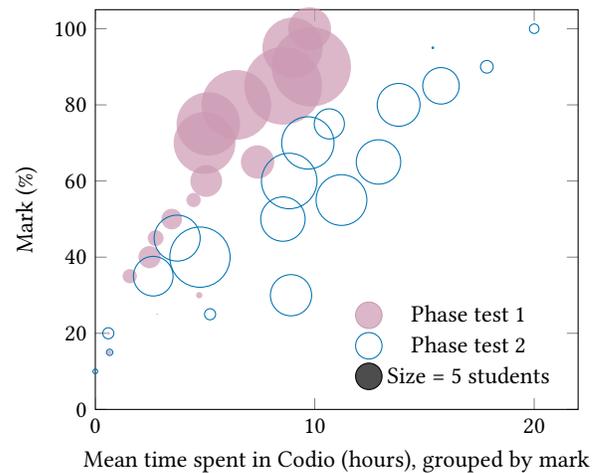
\begin{figure}
	\begin{tikzpicture}
		\begin{axis}[ylabel={Mark (\%)},
					xlabel={Mean time spent in Codio (hours), grouped by mark},
					legend pos=south east,
					legend style={draw=none},
					ymin=0,ymax=105,
					xmin=0,
					xtick distance=10,
					]
					
			\addplot[scatter,
					color=pt1,
					only marks,
					mark size=5,
					fill opacity=0.7,
					scatter src=explicit,
					visualization depends on=\thisrow{pt1count}\as\meta,
					scatter/use mapped color={color=pt1, draw opacity=0},
					scatter/@pre marker code/.append style={
						/tikz/mark size=0.3*\meta}]
					table[x=pt1,y=mark,meta=pt1count]{time_summative.dat};
			\addlegendentry{Phase test 1}
			
			\addplot[scatter,
					color=pt2,
					only marks,
					mark size=5,
					fill opacity=0.0,
					scatter src=explicit,
					visualization depends on=\thisrow{pt2count}\as\meta,
					scatter/use mapped color={color=pt2, draw opacity=1.0},				
					scatter/@pre marker code/.append style={
						/tikz/mark size=0.3*\meta}]
					table[x=pt2,y=mark,meta=pt2count]{time_summative.dat};
			\addlegendentry{Phase test 2}

			\addplot[scatter,
					color=black,
					only marks,
					mark size=5,
					fill opacity=0.7,
					scatter src=explicit,
					visualization depends on=\thisrow{fakecount}\as\meta,
					scatter/use mapped color={color=black, draw opacity=1.0},				
					scatter/@pre marker code/.append style={
						/tikz/mark size=0.3*\meta}]
					table[x=fake,y=mark,meta=fakecount]{time_summative.dat};
			\addlegendentry{Size = 5 students}		
		\end{axis}
		\end{tikzpicture}
	\caption{Relationship of marks to time engaged in Codio.}
	\label{fig:summative_relationship}
	\vskip-12pt
\end{figure}

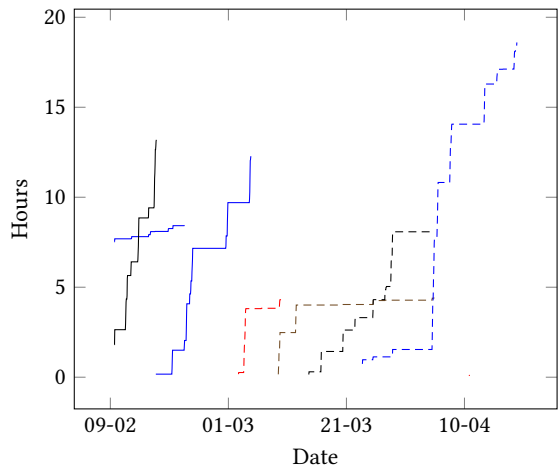
\begin{figure}
\begin{tikzpicture}
	\begin{axis}[
			date coordinates in=x,
			xticklabel style={anchor=near xticklabel}, 
			xticklabel=\day-\month,
			xlabel={Date},
			ylabel={Hours},
			]
		\addplot table[col sep=comma,x=date,y=value,mark=none] {intro.csv};
		\addplot table[col sep=comma,x=date,y=value,mark=none] {testing.csv};
		\addplot table[col sep=comma,x=date,y=value,mark=none] {algorithms.csv};
		\addplot table[col sep=comma,x=date,y=value,mark=none] {searching.csv};
		\addplot table[col sep=comma,x=date,y=value,mark=none] {sql.csv};
		\addplot table[col sep=comma,x=date,y=value,mark=none] {pointers.csv};
		\addplot table[col sep=comma,x=date,y=value,mark=none] {intermed.csv};
		\addplot table[col sep=comma,x=date,y=value,mark=none] {structs.csv};
		\addplot table[col sep=comma,x=date,y=value,mark=none] {sorting.csv};
	\end{axis}
\end{tikzpicture}
\caption{Example of engagement tracking for an individual student. Each line is time spent on a week's Codio activities.}
\label{fig:engage}
\end{figure}

This link from engagement to achievement is well established \cite{trowler2010student} but not always obvious to students.  Now when we make the claim we can back it up with aggregated evidence produced by our own students.  For example, Figure \ref{fig:summative_relationship} shows the relationship between minutes in Codio and marks on relevant assessment for one cohort $-$ the linear relationship between the two apparent.  Further, we can show students their own data and a predicted final mark based on current engagement.  Figure \ref{fig:engage} shows an example of the type of data that could be extracted for an individual student.  
Figures \ref{fig:summative_relationship} and \ref{fig:engage} make clear the strong correlation between Codio engagement and grades.  To be clear, the inclusion of these figures is not to make a claim of causation derived from Codio use, but to demonstrate the type of data that can be extracted for the purposes of convincing students of the importance of engagement!

During the 2017/18 year the Codio data was combined with information from the university's register system and used to send emails to all students.  These emails were automated, but also personalised (templates with blanks for personal data).  They gave specific actionable advice based on individual situation, e.g. if doing well told to keep up the good work; or if evidence of missing knowledge told how to catch up.  The emails made mention of the specific topics of study: possible via the Codio tracking.
%
%
%
%

\subsection{Performance and Satisfaction}

The second author ran the same module in 2016/17 without Codio and then in 2017/18 with Codio. The other changes made between these two years were minimal.  The material that formed the Codio guides was taken from existing pdfs and the tasks from existing lab questions.  The entry criteria onto the degrees did not change significantly and the teaching style was otherwise unchanged.


The summative assessment points (two multiple choice quizzes and a written exam) were similar in style, structure, and difficulty between the two years\footnote{While of course having different questions to maintain integrity.}.  Overall pass rates remained broadly the same.  However, average marks increased significantly.  The average grade on the first quiz increased from 57\% to 63\%, while the grade on the written exam from 51\% from 59\%.  It seems Codio had improved performance for most, but not the very weakest students.

Module satisfaction is measured using an online questionnaire that mimics the UK National Student Survey.  For this module overall satisfaction increased significantly from 82\% to 89\%.   154 students left free text comments that mentioned Codio, with 84\% of those positive. Student feedback is almost unanimously in favour of automated feedback, however, views on Codio itself are more diverse.  While a large majority like the system, some are vocally against. 

Those against are often the students with the most prior programming experience who resent being denied the use of their \ac{IDE} of choice\footnote{It is certainly true that environments such as CodeBlocks and Visual Studio have more features (and a wealth of online tutorial resources) but students who use these systems can lack an understanding of what tasks are being performed for them. \\
For example, we force students in Codio to manually compile their C++ programs at the start deliberately, to ensure they have an understanding of the process before moving on to using automated build tools.}.
We note that in addition to the tasks set in Codio we also set students larger extended projects and problems which can be tackled on any platform the student chooses (such as the ALL projects mentioned in footnote 1).   We tend to use Codio for the basic (and thus most important) tasks that introduce the material.

\section{Challenges}
\label{SEC:Challenges}

Some challenges the authors encountered include the following:
\begin{itemize}
    \item The initial creation of the automated tests requires a large up front investment of time.
    \item Dependency on an online system means that in the unlikely event of a campus internet failure classes come to a halt.
    \item Codio itself has been found to contain occasional bugs.  The Codio support team acted rapidly to fix these.
\end{itemize}
The greatest challenges though, have been in the legal and administrative side of things.  GDPR requirements were an original motivation for using Codio, but the university's changing interpretation of them has created further hurdles, e.g. on the location of hosted data and how student consent is taken.  
Although Codio does have built-in integration for popular \acp{VLE} such as Moodle, Coventry regulations have blocked our use of these, leading to a painful manual signup procedure.

\section{Future Work}
\label{SEC:Future}

We see a number of future developments in this initiative, especially in further use and analysis of the data Codio produces.  Given the success in formative assessment an obvious next step is the automated assessment of summative work.  Whilst this does pose some challenge in assessment design, again, the main barriers to date are university regulations: as a \nth{3} party platform, Coventry currently does not allow the use of Codio for summative assessment.

We are now extending our existing automated summative assessment (multiple choice quizzes in the Moodle \ac{VLE}) to include question types that ask students to write code which is evaluated against unit tests.  For this we are using \texttt{Coderunner} \cite{coderunner}, a free open source Moodle plugin that can run entirely on university hardware.

\section{Summary}

We now extensively use automated feedback and cloud VMs in the programming curriculum at Coventry University, with positive results and further developments underway.  A message for colleagues with similar plans is that difficulties may include not only upfront development costs but also legal and institutional regulations.

\subsection*{ACKNOWLEDGEMENTS}

The authors thank their colleagues who teach in these classes, and acknowledge the support of The Institute of Coding.

\bibliographystyle{ACM-Reference-Format}
\bibliography{CE19}

\end{document}